\documentclass[prl,showpacs,amsfonts,amssymb,amsmath]{revtex4}
\usepackage{graphicx}

\begin{document}

\title{Finite time collapse of N classical fields described by coupled nonlinear Schr\"odinger equations}

\author{D. C. Roberts$^1$ and A. C. Newell$^2$} \affiliation{$^1$Laboratoire de Physique Statistique de l'Ecole Normale Sup\'erieure, Paris, France \\ $^2$Department of Mathematics, University of Arizona, Tucson, AZ} 
\begin{abstract}
We prove the finite-time collapse of a system of N classical fields, which are described by N coupled nonlinear Schr\"odinger equations.  We derive the conditions under which all of the fields experiences this finite-time collapse.  Finally, for two-dimensional systems, we derive constraints on the number of particles associated with each field that are necessary to prevent collapse.
\end{abstract}

\maketitle
 
The nonlinear Schr\"odinger equation (NLSE) arises in many diverse areas of physics, such as nonlinear optics, gravity waves in deep water and, recently, trapped dilute Bose-Einstein gases where it has been important in explaining many of the static and dynamic properties of dilute Bose-Einstein condensates (BECs) \cite{nlse,pit}.  It has been proven that, under certain conditions, the NLSE exhibits finite-time collapse, which is the occurrence of a finite time singularity in the field.  In this paper, we generalize this proof to show that finite-time collapse occurs in N-coupled nonlinear Schr\"odinger equations and we identify the parameter regime and initial conditions required for this collapse.  We motivate the work of this paper by referring to the state-of-the-art in experimental research on BECs.

The NLSE was shown, in the context of Langmuir waves in plasma physics, to undergo finite-time collapse for systems with a negative coupling constant\cite{zakharov}.  Finite-time collapse was indicated by the absence of variance in the field described by the NLSE at a finite time.  Proof of this finite-time collapse of the "focusing NLSE" has been important for understanding the stability of trapped single-component BECs.  In particular, this proof implies that a condensate formed by atoms with a negative coupling constant, i.e. atoms that attract each other, would in theory develop a singular density; in reality, three-body collisions disallow the development of an infinite density \cite{3body}.  (In experimental setups involving a small number of attractive atoms, however, an external potential can be used to stabilize the dilute BEC system even though the coupling constant is negative.)

We move now to the generalization of this proof of the focusing NLSE's finite-time collapse to that of N coupled NLSEs describing a system of coupled classical fields.  In generalizing the proof, we shall also determine the range of initial conditions and the parameter regime over which all of these coupled fields exhibit instability.  The present exercise is of particular importance given the shift in focus of BEC experiments from single-component condensates to condensates of miscible mixtures.  Current experiments are able to achieve two component condensates (see for example \cite{jila, diffatoms}), and the prospects for realizing condensates of many more components are good given, for instance, the recent achievement of condensed Ytterbium \cite{yt}, which has five bosonic isotopes.    

Coupled NLSEs have been successfully employed to model many dynamic and static properties of binary mixtures in recent dilute Bose-Einstein condensate experiments (see for example \cite{pit, coupled}).   As a generalization of this approach for multicomponent systems, let us consider the following N coupled NLSEs describing the behavior of the classical fields $\psi_{i}$ of the system:
\begin{equation}
\label{nlse}
i\partial_t \psi_{i}({\bf r},t)= - \vec{\nabla}^2 \psi_i({\bf r},t)+\sum_{j=1}^N g_{ij}|\psi_{j}({\bf r},t)|^2 \psi_{i}({\bf r},t)
\end{equation}
where $g_{ij}$ are the coupling constants (which, in the BEC context, can be written in terms of the relevant two-body scattering length).   For simplicity we will assume $g_{ij}=g_{ji}$ and no external potentials.

To find the conditions that result in finite-time collapse of all classical fields, we examine the sum of the variances of the fields in the system, which can be expressed as
\begin{equation}
V(t)=\sum_{i=1}^N V_i(t)=\sum_{i=1}^N \int d {\bf r} r^2 |\psi_{i}({\bf r},t)|^2.
\end{equation}
When $V(t)=0$, since the variance of each given field, $V_i(t)$ is necessarily positive, all of the coupled classical fields must have a variance of zero, i.e. all fields become singular.  

To prove that $V(t) \rightarrow 0$ in a finite time, using eq. \ref{nlse}, we calculate the variance for a given field (through successive integration by parts) 
\begin{equation}
\partial_t^2 V_i(t)= \int d {\bf r} \left[ r^{d-1} 8 |\partial_r \psi_i(r,t)|^2+2 d g_{ii} |\psi_i(r,t)|^4-4 r^d \sum_{i \ne j}^N g_{ij}|\psi_i(r,t)|^2  \partial_r |\psi_j(r,t)|^2 \right]
\end{equation}
where, for simplicity, we have ignored the angular variables and, for completeness, we have kept the dimension of space $d$ as a variable. Simplifying further using integration by parts, we arrive at the crucial result (which was not at all obvious for the multicomponent system) for the proof of the finite time collapse for coupled NLSEs:
\begin{equation}
\label{var}
\partial_t^2 V(t) = 8 H +4 (d-2) U(t).
\end{equation}   
  Here the Hamiltonian, which is a constant of motion of this system, is given by
\begin{equation}
H=T(t)+U(t) 
\end{equation}
where $T(t)$ corresponds to the kinetic energy of the system and is given by 
\begin{equation}
T(t)= \sum^{N}_{i=1} \int d^3 {\bf r}|\vec{\nabla} \psi_i({\bf r},t)|^2;
\end{equation}
and $U(t)$ corresponds to the potential energy of the system and is given by  
\begin{equation}
U(t)=\frac{1}{2} \int d^3 {\bf r} g_{ij} |\psi_i({\bf r},t)|^2  |\psi_j({\bf r},t)|^2,
\end{equation} 
summing over repeated indices.

Assuming $d=3$, we must impose the condition $U(t=0) \le 0$ in order to prove finite-time collapse using the variance method presented in this paper.  This implies that the matrix $g_{ij}$ must be negative semi-definite \cite{fnr}.     If $d=2$, there is no need to assume $U(t)$ is negative.   Then, following the logic presented in \cite{sulem} for the single NLSE, we impose on eq. (\ref{var}) the condition that $g_{ij}$ be negative semi-definite for $d=3$ (no such condition is imposed for $d=2$) to arrive at 
\begin{equation}
\partial_t^2 V(t) \le 8 H.
\end{equation}
Since $H$ is a constant of motion, we can integrate this equation to obtain 
\begin{equation}
\label{vart}
V(t) \le 4H t^2 +V'(0)t+V(0).
\end{equation}
  The right-hand side will go to zero in a finite time - which, as argued above, implies that every one of the fields $\psi_{i}$ must become singular within a finite time - if any one of the following three sets of conditions is true:
\begin{enumerate}
\item $H<0$ 
\item $H=0$ and $V'(0)<0$
\item $H>0$ and $V'(0)<-4 \sqrt{H V(0)}$
\end{enumerate}
Note that the time given in eq. (\ref{vart}) is only an upper bound on the collapse time of the entire system.

The fact that the final term in eq. (\ref{var}) is proportional to $U(t)$, which was not expected a priori, greatly simplifies the proof of finite time collapse of classical fields whose behavior is determined by coupled NLSEs.  This leads to the simple conclusion that if the potential energy, $U(t)$, is sufficiently negative to overcome the kinetic energy, $T(t)$, collapse will occur.  

As an example, let us assume initially that every one of the fields is uniform, i.e. $T(t=0)=0$.  In this case, a sufficient condition for finite-time collapse of all fields is that the matrix $g_{ij}$ be negative semi-definite, which assures $U(0)<0$.  Therefore, as the sum of $T(0)$ and $U(0)$, $H(0)<0$, which satisfies the first set of conditions above.  

If $d=2$, one can derive strict contraints on the number of particles in each component given by $N_i(t)=\int d {\bf r} |\psi_i({\bf r},t)|^2$ in order for the coupled classical fields not to collapse.  For clarity, let us first derive the analogous constraint on the number of particles for the collapse of a single field obeying the NLSE.

The proof relies on the Gagliardo-Nirenberg inequality given by (note that the discussion below assumes $d=2$)
\begin{equation}
\label{GN}
\int d {\bf r} |\psi({\bf r},t)|^4  \le C N(t) T(t)
\end{equation}
where C is a positive constant determined by the appropriate Euler-Lagrange equations \cite{constant}.  Note that we have temporarily removed the subscript as we are dealing here with only one non-coupled component.   Using this inequality, one can show that 
\begin{equation}
H \ge \int d {\bf r} |\psi({\bf r},t)|^4 \left[ \frac{1}{C N(t)}+\frac{g}{2} \right].
\end{equation}
Therefore, if $g \ge 0$, or if $N(t)<-2/g C$ and $g<0$, then the potential energy $U(t)$ is bounded by a finite amount, which implies that the kinetic energy is also bounded, and hence collapse is not possible. 

One can generalize this argument to the case considered in this paper of $N$ classical fields governed by coupled NLSEs given by eq. (\ref{nlse}) for $d=2$.  Using the inequality given by eq. (\ref{GN}), one can show that 
\begin{equation}
H(t) \ge \int d {\bf r} \Lambda_{ij} |\psi_i({\bf r},t)|^2  |\psi_j({\bf r},t)|^2
\end{equation}
where $\Lambda_{ij}=\delta_{ij}/C N_i(t)+g_{ij}/2$ and $\delta_{ij}=1$ when $i=j$ and zero otherwise.   If $\Lambda_{ij}$ is positive definite \cite{negdef}, then there will be no collapse as the potential energy is bounded by a finite amount.   This condition, assuming $g_{ij}$ is given, puts constraints on the number of particles in each coupled field to ensure there is no collapse in $d=2$, analogous to the single-component case considered above.  Note that if $\Lambda_{ij}$ is not positive definite, then this analysis does not give information about whether or not collapse will occur. 

The authors are grateful to Y. Pomeau for many stimulating discussions.  DCR acknowledges financial support from the Marie Curie Fellowship.


\begin{thebibliography}{9}
\bibitem{constant} The best value of $C$ is given by $2/|R|^2$ where $R$ is the positive solution of $\Delta R -R+R^3=0$ \cite{sulem}.
\bibitem{fnr} Interestingly, the condition for thermodynamic and dynamic stability for an N-component miscible homogeneous condensate is that the matrix $g_{ij}$ be positive semi-definte \cite{roberts}.
\bibitem{negdef} Positive definite implies that all eigenvalues of a given matrix must be positive real numbers.
\end{thebibliography}
\end{document}